\documentclass[acmsmall,screen]{acmart}

\usepackage{booktabs} 
\usepackage{algorithm}
\usepackage{algpseudocode}
\usepackage{multirow}
\usepackage{tabularx}
\usepackage{array}
\usepackage{subcaption}
\usepackage{xfrac}
\usepackage[normalem]{ulem}

\setcopyright{cc}
\setcctype{by}
\acmJournal{PACMCGIT}
\acmYear{2026} \acmVolume{9} \acmNumber{4} \acmArticle{56}
\acmMonth{7} \acmDOI{10.1145/3820023}
\newif\ifdrafting
\draftingfalse 

\algnewcommand{\ParFor}[1]{\textbf{Parallel for all} #1 \textbf{do}}

\newif\ifredline
\redlinefalse  

\ifredline
  \newcommand{\redline}[1]{\textcolor{red}{#1}}
\else
  \newcommand{\redline}[1]{#1}
\fi

\ifdrafting
     \usepackage[svgnames,x11names,table]{xcolor}
     \newcommand{\jk}[1]{\textcolor{magenta}{[Jonghyun: #1]}}
     \newcommand{\dl}[1]{\textcolor{orange}{[Dave: #1]}}
     \newcommand{\cs}[1]{\textcolor{purple}{[Cheng: #1]}}
     \newcommand{\jj}[1]{\textcolor{DarkBlue}{[Jaehyun: #1]}}
     \newcommand{\sd}[1]{\textcolor{blue}{[Shalini: #1]}}
     \newcommand{\ar}[1]{\textcolor{green}{[Andrew: #1]}}
     \newcommand{\ms}[1]{\textcolor{red}{[Michael: #1]}}
     \newcommand{\wb}[1]{\textcolor{red}{[Wil: #1]}}
 \else
     \newcommand{\jk}[1]{}
     \newcommand{\dl}[1]{}
     \newcommand{\cs}[1]{}
     \newcommand{\jj}[1]{}
     \newcommand{\sd}[1]{}
     \newcommand{\ar}[1]{}
     \newcommand{\ms}[1]{}
     \newcommand{\wb}[1]{}
 \fi

\AtBeginDocument{%
  \providecommand\BibTeX{{%
    \normalfont B\kern-0.5em{\scshape i\kern-0.25em b}\kern-0.8em\TeX}}}

\begin{document}

\title{Real-time 3D Visualization of Radiance Fields on Light Field Displays}

\begin{teaserfigure}
\centering
\includegraphics[width=\textwidth]{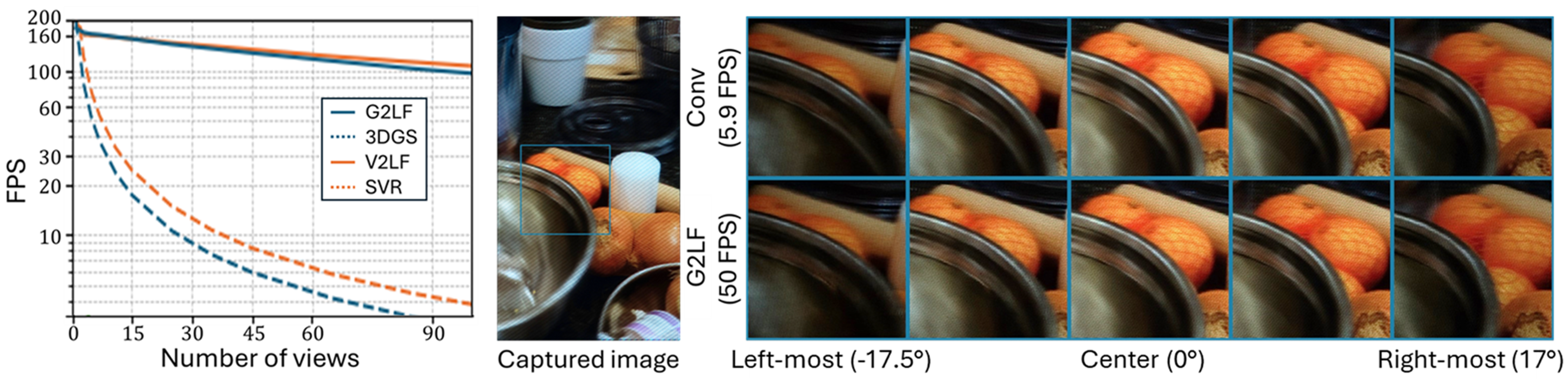}
\caption{We present a unified framework for real-time radiance field rendering on light field displays, supporting explicit representations including 3D Gaussians (G2LF) and sparse voxels (V2LF) via single-pass plane sweeping over intermediate multi-plane images (MPI). (Left) Render time comparison across view counts on the MipNeRF-360 dataset \cite{barron2022mip}, measured at 512p on a NVIDIA RTX 5090. G2LF and V2LF (solid lines) maintain real-time performance (>60 FPS) beyond 90 views. (Right) Captured results of a 45-view rendered light field shown on a Looking Glass Go display, recorded with a motorized rotational stage, demonstrate high-quality 3D reconstruction across viewpoints. Note that the performance plot uses $N_{\text{chunk}}=64$, whereas the captured light field on the right uses $N_{\text{chunk}}=512$ for higher visual quality (see Sec.~\ref{sec:sweeping}).}
\Description{On the left, a line chart plots rendering speed (frames per second) against the number of rendered views for four methods (G2LF, 3DGS, V2LF, SVR); the proposed G2LF and V2LF stay above 60 FPS as views increase while the baselines fall off sharply. On the right, a scene shown on a Looking Glass light field display is photographed from left, center, and right viewpoints, comparing conventional rendering with our G2LF.}
\label{fig:teaser}
\end{teaserfigure}

\author{Jonghyun Kim}
\orcid{0000-0002-1197-368X}
\authornote{contributed equally to this research}
\affiliation{
  \institution{NVIDIA}
  \city{Santa Clara}   
  \state{CA}
  \country{USA}
}
\email{jonghyunk@nvidia.com}   

\author{Cheng Sun}
\orcid{0000-0001-8910-4524}
\authornotemark[1]
\affiliation{
  \institution{NVIDIA}
  \city{Santa Clara}   
  \state{CA}
  \country{USA}
}
\email{chengs@nvidia.com}       

\author{Michael Stengel}
\orcid{0009-0008-6234-936X}
\authornotemark[1]
\affiliation{
  \institution{NVIDIA}
  \city{Santa Clara}   
  \state{CA}
  \country{USA}
}
\email{mstengel@nvidia.com}     

\author{Matthew Chan}
\orcid{0009-0009-1704-1013}
\affiliation{
  \institution{NVIDIA}
  \city{Santa Clara}   
  \state{CA}
  \country{USA}
}
\email{matchan@nvidia.com}      

\author{Andrew Russell}
\orcid{0009-0002-0820-115X}
\affiliation{
  \institution{NVIDIA}
  \city{Santa Clara}   
  \state{CA}
  \country{USA}
}
\email{arussell@nvidia.com}     

\author{Jaehyun Jung}
\orcid{0000-0002-5709-6045}
\affiliation{
  \institution{NVIDIA}
  \city{Santa Clara}   
  \state{CA}
  \country{USA}
}
\email{jaehyunj@nvidia.com}     

\author{Wil Braithwaite}
\orcid{0009-0005-1741-0748}
\affiliation{
  \institution{NVIDIA}
  \city{Santa Clara}   
  \state{CA}
  \country{USA}
}
\email{wbraithwaite@nvidia.com} 

\author{David Luebke}
\orcid{0000-0002-8206-5785}
\affiliation{
  \institution{NVIDIA}
  \city{Santa Clara}   
  \state{CA}
  \country{USA}
}
\email{dluebke@nvidia.com}      

\author{Shalini De Mello}
\orcid{0009-0009-0213-2860}
\affiliation{
  \institution{NVIDIA}
  \city{Santa Clara}   
  \state{CA}
  \country{USA}
}
\email{shalinig@nvidia.com}     

\begin{abstract}
Radiance fields, including their recent efficient forms such as 3D Gaussian Splatting and Sparse Voxels, have revolutionized photorealistic 3D scene visualization by enabling high-fidelity reconstruction of complex environments, making them a natural match for light field displays. However, integrating these technologies presents significant computational challenges, as light field displays require many high-resolution renderings from slightly shifted viewpoints, while radiance fields rely on computationally intensive volume rendering, which is intractable to achieve real-time speeds even with efficient scene representations. In this paper, we propose a unified and efficient framework for real-time radiance field rendering on light field displays. Rather than re-rendering each view independently, our method converts the input radiance field into shared intermediate sweeping planes that can be efficiently composited into dense light-field views in a single pass. \redline{Our method prioritizes shared, non-directional plane caching for real-time performance, trading fine view-dependent color effects for a modest increase in intermediate memory usage.} Our framework generalizes across different scene representations without retraining and avoids repeated computation across views. We further demonstrate a real-time interactive application on a Looking Glass display, achieving 200+ FPS at 512p across 45 rendered views and enabling seamless, immersive 3D interactive viewing experiences. On standard benchmarks, our method achieves up to 22× speedup compared to independently rendering each view, while largely preserving image quality.
\end{abstract}

%
%

\ccsdesc[500]{Computing methodologies~Rendering}
\ccsdesc[300]{Computing methodologies~Appearance and texture representations}
\ccsdesc[300]{Computing methodologies~Shared memory algorithms}



\maketitle

\section{Introduction}

Recent advancements in radiance fields have significantly improved both the quantity and quality of 3D content. Radiance fields represent 3D scenes by encoding density and color values across spatial coordinates and view directions, enabling photorealistic reconstructions of complex scenes. Neural Radiance Fields (NeRF) have enabled continuous view synthesis from sparse input images, making it possible to reconstruct complex 3D scenes with high precision \cite{mildenhall2021nerf, muller2022instant}. Radiance fields now encompass a variety of representations for improving rendering efficiency for real-time applications, including rasterization-based methods \cite{kerbl20233d, sun2024sparse} and explicit scene structures \cite{takikawa2021neural, sun2022direct} that avoid dense sampling.

Like other 3D content, radiance fields are most effectively visualized using 3D displays. Their ability to represent complex 3D structures aligns naturally with the capabilities of light field displays, which optically reconstruct the light rays of 3D scenes. Recent commercially available light field displays offer high spatial and angular resolution, enabling immersive 3D visualizations (\cite{leia}, \cite{sony}, \cite{lookingglass}). These displays provide binocular disparity and motion parallax, leveraging human depth perception to allow users to naturally perceive 3D structures. However, the integration of neural radiance fields with light field displays, introduces significant computational challenges. First, light field displays require high-resolution rendering from many (45+) slightly shifted viewpoints. Second, generating this many views from radiance fields, even with fast rasterization-based methods like 3D Gaussian Splatting \cite{kerbl20233d} and Sparse Voxels \cite{sun2024sparse} remains computationally intractable for real-time applications such as interactive 3D/4D content viewing.

\emph{Light field displays} fundamentally face substantial computational overhead in rendering due to their unique optical design. Unlike conventional single-view 2D displays, they require the generation of multiple perspective views to reconstruct the full light field, necessitating a dense array of rays projected at precise angles. This significantly increases the computational burden compared to single-view 2D displays. Furthermore, precise optical alignment between the display panel and the lens array is critical; even minor angular or spatial misalignment during manufacturing can lead to incorrect ray-to-subpixel mappings, degrading visual quality (see Section \ref{sec:3d} for details). Correcting these errors requires per-device calibration and often a larger number of rendered views, which further increases the rendering cost.

\emph{Radiance fields} face inherent computational challenges, particularly due to their reliance on volumetric rendering, where density and color values must be evaluated along each ray. To reduce these costs, recent approaches employ explicit/hybrid data structures and rasterization-based pipelines to accelerate rendering, such as 3D Gaussian Splatting (3DGS) \cite{kerbl20233d} and Sparse Voxels \cite{sun2024sparse}. However, when targeting light field displays, these methods still require rendering many slightly shifted viewpoints, making repeated sampling inefficient and potentially redundant. Addressing this redundancy is essential to enabling real-time radiance field rendering for light field applications.

In this paper, we present a unified framework for real-time radiance field rendering on light field displays. Our framework supports a wide range of explicit radiance field representations—including 3DGS and Sparse Voxels—within a unified rendering architecture. The core of our method is a single-pass of plane sweeping over multi-plane images (MPI) rendered from the input radiance field that enables efficient view synthesis for multiple novel views while \redline{largely} preserving image quality. We implement this framework in two variants: 3DGS-to-light-field (G2LF) and sparse-voxels-to-light-field (V2LF), corresponding to different input representations. On standard benchmarks, our method achieves up to 22$\times$ speedup compared to independently rendering each view, while preserving \redline{high} image quality. \redline{This efficiency comes from rendering and caching shared intermediate planes that are reused across nearby quilt views. The resulting trade-off is that the representation favors view-independent color caching over fine view-dependent effects, while requiring additional memory for the intermediate planes.} We further demonstrate the effectiveness of our approach through an interactive application running on a commercial light field display, enabling smooth and immersive real-time 3D visualization of radiance fields. Our OpenGL-based interactive demo achieves 228 FPS (22$\times$ faster compared to native quilt rendering, see section \ref{sec:interactive}) for 45-view, 512$\times$910 light field images on the \textsc{bicycle} scene using a single NVIDIA RTX 5090 graphics card (see supplementary video).

The contributions of this paper are as follows: 
\begin{itemize} 
\item We propose a unified and efficient multi-view rendering framework for real-time rendering of radiance fields on light field displays, supporting multiple explicit (3DGS and Sparse Voxels) representations without retraining them.
\item Our method achieves real-time performance (>60 FPS for 90+ views, up to 228 FPS for 45 views) by minimizing redundant sampling through a single-pass plane sweeping strategy.
\item We develop an OpenGL-based interactive 3DGS renderer for a commercial light field display, enabling real-time 3D visualization and achieving a 3.6$\times$ speedup over our Python/CUDA baseline.
\end{itemize}

\section{Related Work}
\label{sec2}

\paragraph{Radiance field representations}
NeRF introduced continuous volumetric scene encoding for novel view synthesis~\cite{mildenhall2021nerf}, prompting numerous efforts to accelerate rendering. Techniques for it include hash-based feature grids~\cite{muller2022instant}, space decomposition~\cite{Reiser2021_KiloNeRF, Chen2022_TensoRF}, and hierarchical training frameworks~\cite{barron2022mip, tancik2023nerfstudio}. More recently, explicit and hybrid representations such as 3DGS~\cite{kerbl20233d} and voxel-based rasterization~\cite{takikawa2021neural, sun2022direct, sun2024sparse} have enabled real-time rendering through rasterization-friendly pipelines. While these methods reduce the cost of rendering a single view, they do not directly address the inefficiency of rendering many closely related views, as required by multi-view displays and remain prohibitively expensive for rendering many views simultaneously. 

\paragraph{Depth-guided view synthesis} 
Classical methods such as depth-image-based rendering (DIBR)~\cite{Fehn2004_DIBR3DTV}, layered depth images (LDI)~\cite{shade1998layered, Broxton2020_LFVideo}, multi-plane images (MPI)~\cite{zhou2018stereo}, \redline{and related homography-based view-synthesis methods} synthesize novel views by projecting input images, often with depth information, onto proxy geometry or discrete depth layers. \redline{Wizadwongsa et al. propose a fast MPI method with a custom data representation that separates training and render path for specific MPIs containing view-dependent effects from those providing diffuse information~\cite{wizadwongsa2021nex}.}
While our method shares structural similarities with these approaches, it differs in both construction and use: rather than estimating a custom layered representation \redline{tailored to a specific display} from input images, we render \redline{MPIs for dense light-field view synthesis directly from an already reconstructed 3D radiance field. This architecture enables broad compatibility for any explicit and increasingly powerful radiance field representation
and allows efficiently sharing rendered information across many closely spaced light-field views, allowing each view to be generated by inexpensive plane reprojection and accumulation instead of repeatedly rendering the original radiance field.}

\paragraph{Multi-view rendering}
Techniques such as multi-view point splatting~\cite{Hubner2006_MVPointSplat}, single-pass multi-view rendering~\cite{Hubner2007_SinglePassMV}, and unstructured lumigraph rendering~\cite{Buehler2001_UnstructuredLumigraph}, \redline{dynamically reparameterized light fields on focal surfaces~\cite{isaksen2000dynamically},} as well as hardware-level instancing methods~\cite{NVIDIA_VRWORKS, unterguggenberger2020fast}, reduce rendering cost across nearby views. However, these methods are tailored to traditional 3D content, such as meshes or point clouds, and are difficult to apply to radiance fields, where sharing computation across nearby views remains challenging. Our method addresses this by providing a unified solution that supports both implicit and explicit representations, enabling single-pass rendering of high-quality multi-view outputs.

\redline{Adaptive projection tailored to the light field display type can reduce the pixel waste resulting from interleaving for all rasterized content but has been shown to introduce subsampling artifacts without custom anti-aliasing strategies~\cite{fink2023efficient}. This line of optimization is orthogonal to our strategy and could potentially be coupled with our method for further performance optimization.}

\paragraph{Multi-view radiance field rendering}
Rendering multiple views from radiance fields is computationally intensive due to high redundancy between adjacent viewpoints. Existing methods typically render each view independently, leading to inefficiencies in multi-view or light field applications~\cite{rabia2024orthoscopic, stengel2023ai, tran2024voodoo}. Volume rendering with precomputed ray-to-subpixel mappings~\cite{chen2022fast, ji2025text, kim20263d} reduces this overhead but requires per-device calibration for accurate rendering. 
\redline{DirectL~\cite{yang2024directl} also targets efficient light-field radiance-field rendering through calibrated ray mappings, but relies on ray-tracing-style sampling, which is less suited to rasterization-based explicit representations.}
\redline{In contrast, our approach is designed for rasterization-based explicit radiance fields: we rasterize each primitive once into shared MPI-like sweeping planes and reuse the cached planes for dense quilt generation. This minimizes redundant sampling across slightly shifted views while providing a unified pipeline for unmodified explicit radiance-field representations.}
\section{Light Field 3D Display}\label{sec:3d}

\begin{figure}[h]
\centering
\includegraphics[width=0.8\linewidth]{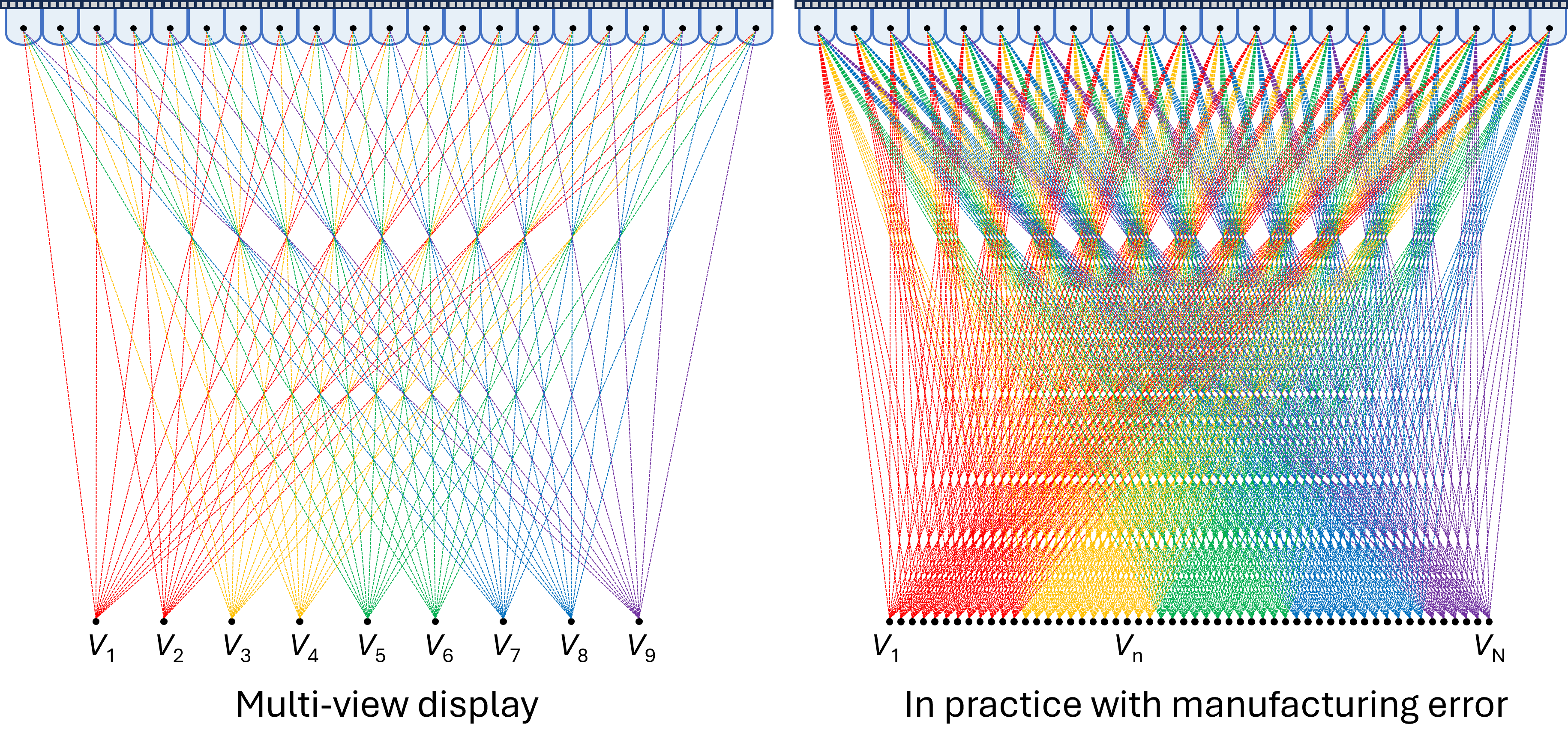}
\caption{Ray diagrams of light field displays as viewed from the top. Left: an ideal imperfection-free multi-view display (9-view); Right: a multi-view display with a small manufacturing misalignment ($0.00884^\circ$) in the lens array. Even with the same nominal 9-view layout, such a small misalignment changes the ray-to-subpixel mapping and breaks the intended view separation. \redline{Because calibrated ray directions can vary across panel pixels, more rendered views may be required to densely sample the corrected view mapping.} In practice, these imperfections necessitate per-device calibration, which can further increase rendering cost. Here, a 1080p liquid crystal display paired with a non-slanted lens array is assumed for simplicity.}
\Description{Top-down ray diagrams of a multi-view light field display. Left: an ideal nine-view display where rays from each microlens converge cleanly to separate viewpoints. Right: the same display with a tiny lens-array misalignment (0.00884 degrees) that scrambles the ray-to-subpixel mapping and breaks the intended view separation.}
\label{fig:ray}
\end{figure}

Multi-view displays are designed to deliver binocular disparities at different viewpoints, providing motion parallax cues for viewer(s). The light rays from each microlens converge to specific viewpoints at the viewing distance  (Fig. \ref{fig:ray}, left).
Such view separation is commonly achieved through optical elements, including parallax barriers \cite{lanman2010content}, lenticular lenses \cite{ives1931optical}, or slanted lenticular lenses \cite{van1999image}, attached to the display panel. However, each of these different rendered views that are delivered to the viewer via a light field display must all be rendered from the original 3D scene and provided together to the display. The rendering process for these multiple views involves placing perspective cameras in parallel at the designated viewing distance from the 3D scene and rendering from them. 
After rendering all views, \redline{the rendered perspective images are arranged into a 2D view atlas, often called a ``quilt'' in light field display systems. The quilt is then interleaved into the final panel image according to the display's optical configuration.}

Recent commercially available light field 3D displays often increase angular resolution to achieve smoother motion parallax (i.e., they increase the number of viewpoints). However, this design choice imposes a significant computational burden as significantly more views must be rendered.
Additionally, manufacturing imperfections, such as angular and spatial misalignments between the lens array and display panel, further increase computational cost. Even a small angular misalignment, such as a half-subpixel offset (only 0.00884$^\circ$ for a 1080p LCD), can scramble all viewpoints (Fig. \ref{fig:ray}, right) \cite{kim2015calibration}. Correcting these errors requires a per-device calibration process to estimate accurate view-to-subpixel mappings, ensuring geometrically correct 3D reconstruction. Although effective, this calibration further increases the number of rendered views and amplifies computational demands. As a result, modern \redline{horizontal motion-parallax} light field displays often require 45 or more rendered views, posing significant challenges for real-time, interactive 3D rendering, particularly when dealing with radiance field content.

Recent multi-view rendering methods reduce computation by leveraging ray-to-subpixel mappings for light field displays, but still rely on per-device calibration~\cite{yang2024directl, chen2022fast, ji2025text}. To address this limitation, we propose a unified framework for radiance field rendering on light field displays. Our method reduces computational redundancy by minimizing repeated sampling across minimally shifted viewpoints, while maintaining overall perceived image quality, and is compatible with various radiance field representations. Detailed algorithmic descriptions follow in the next section.

\newcommand{\vn}{V}
\newcommand{\vx}{V_\mathrm{x}}
\newcommand{\vy}{V_\mathrm{y}}
\newcommand{\vk}{V_\mathrm{k}}
\newcommand{\distfocal}{D_\mathrm{focal}}
\newcommand{\fov}{\theta}
\newcommand{\fovx}{\fov_\mathrm{x}}
\newcommand{\fovy}{\fov_\mathrm{y}}
\newcommand{\viewfov}{\phi}
\newcommand{\viewfovx}{\viewfov_\mathrm{x}}
\newcommand{\viewfovy}{\viewfov_\mathrm{y}}
\newcommand{\distforward}{D_\mathrm{forward}}
\newcommand{\distshift}{D_\mathrm{shift}}
\newcommand{\planescale}{P_\mathrm{s}}
\newcommand{\rgbplanes}{\mathbf{C}}
\newcommand{\Tplanes}{\mathbf{T}}
\newcommand{\quilts}{\mathbf{Q}}
\newcommand{\Nchunk}{N_\mathrm{chunk}}
\newcommand{\Nx}{N_\mathrm{x}}
\newcommand{\Ny}{N_\mathrm{y}}
\newcommand{\off}{\Delta}
\newcommand{\offx}{\off_x}
\newcommand{\offy}{\off_y}
\newcommand{\cx}{c_x}
\newcommand{\cy}{c_y}
\newcommand{\ck}{c_k}
\newcommand{\distplanes}{\mathbf{d}}

\section{Radiance Fields to Light Field}

We first give an overview of our efficient radiance fields to light fields rendering pipeline in Sec.~\ref{sec:overview}, and introduce the shared components in Sec.~\ref{sec:sweeping} and Sec.~\ref{sec:swizzling}. Later, we adapt our algorithm to various state-of-the-art efficient 3D representations commonly used in radiance field reconstruction and novel-view synthesis. The representation-specific adaptations are described in Sec.~\ref{sec:g2lf} and \ref{sec:v2lf}.

\subsection{Rendering pipeline overview} \label{sec:overview}
Given a radiance field and a 3D display viewing setup, our goal is to render a set of $\vn = (\vx \times \vy)$ perspective viewpoints derived from this setup, called ``light field quilts" as illustrated in Fig.~\ref{fig:ray}.
A conventional method applies a radiance field renderer $\vn$ times, separately. However, such a simple approach can slow down rendering up to $\vn$ times, which prevents interactive experiences even for a $200$ FPS single-frame renderer for a common $\vn{=}45$ light field display.

Our strategy is to reduce computation across the multiple rendered views by approximating the volume visible to the 3D display by a series of forward-sweeping planes.
Each plane represents a disjoint frustum of the original volume.
We can then composite the sweeping planes into the $\vn$ quilt views via an operation that we call \textit{swizzle} blending, instead of repeatedly traversing through the original 3D representation to render each of the multiple views, separately.
A single pixel lookup on the planes corresponds to the rendering result of a ray segment from the original radiance field, where the latter is significantly more expensive to compute.
Thus the overall rendering time can be largely reduced as long as we can also render the sweeping planes efficiently.
Next, we introduce the algorithm for sweeping planes in Sec.~\ref{sec:sweeping} and \textit{swizzle} blending in Sec.~\ref{sec:swizzling}.

\begin{figure}[t]
\centering
\includegraphics[width=\linewidth]{figures/Fig_sweeping_planes_revision.png}
\caption{
An illustration of the 3D display parameterization and the sweeping planes.
Left: the focal-plane setup, where $\fov$ is the base camera's field of view and $\viewfov$ is the 3D display's viewing angle.
Center: the reference camera is positioned to cover the entire visible volume behind the focal plane, and the sweeping planes are placed at different depths.
The purple region denotes a near-field volume excluded from the reference camera frustum, which can be reduced in practice by shifting the reference camera backward by $\distshift$.
\redline{Right: after rasterizing scene primitives into the sweeping planes, swizzle blending samples the corresponding plane locations for each quilt-view ray and alpha-composites the sampled colors and transmittances into the final light-field quilt. The sampling coordinate is defined in Eq.~\eqref{eq:u_prime}.}
}
\Description{A three-panel schematic of the sweeping-plane method. Left: the focal-plane setup with the base-camera field of view and the display viewing angle. Center: a reference camera covering the visible volume behind the focal plane, with sweeping planes placed at increasing depths and a purple region marking the excluded near-field. Right: swizzle blending samples these planes along each quilt-view ray and alpha-composites them into the final light-field quilt.}
\label{fig:sweeping_planes}
\end{figure}

\subsection{Sweeping planes rendering} \label{sec:sweeping}
Let's first define the volume of interest of a 3D display setup, which is illustrated in Fig.~\ref{fig:sweeping_planes}'s left panel.
We can imagine the 3D display as a ``window'' (called the focal plane) for viewers to look into the virtual 3D world.
The center location of the focal plane is defined by a \emph{base} camera and a camera-to-plane distance $\distfocal$.
The size of the focal plane is derived from the base camera's field of view $\fovx, \fovy$.
The viewing angles of the 3D display are defined by $\viewfovx, \viewfovy$.

To create a series of sweeping planes, we use a perspective \emph{reference} camera that is adjusted to cover the maximum viewing angle, as shown in Fig.~\ref{fig:sweeping_planes}'s center panel.
For the horizontal $x$ and the vertical $y$ axes, we compute a forward shift $\distforward$ along the $z$-axis required to cover the display's visible volume and take the maximum over the two axes:
\begin{equation}
    \distforward = \max_{k\in\{x,y\}} \distfocal \cdot \frac{\tan(0.5\viewfov_k)}{\tan(0.5\viewfov_k) + \tan(0.5\fov_k)} ~,
\end{equation}
and set the field of view of the reference camera (blue camera in Fig.~\ref{fig:sweeping_planes}) as:
\begin{equation}
    \fov_{k\in\{x,y\}}' = 2 \cdot \arctan(\frac{\distfocal \cdot \tan(0.5\fov_k)}{\distfocal - \distforward}) ~.
\label{eq:theta_prime}
\end{equation}
By construction, the reference camera covers the visible volume behind the focal plane. 
However, some near-field regions between the outermost quilt-view cameras ($\vn$) and the focal plane are excluded (purple region in Fig.~\ref{fig:sweeping_planes}). 
As a simple workaround, we introduce a hyperparameter $\distshift$ that shifts the reference camera backward, i.e., away from the focal plane, such that
$\distforward \leftarrow \distforward - \distshift$. 
This adjustment expands the covered region toward the near-field and reduces the amount of excluded volume in practice.

Finally, the sweeping planes are generated by rendering the scene primitives' chunks or volume chunks into the perspective reference camera. Intuitively, a \textit{chunk} in our context implies the 3D volume between two discrete sweeping plane locations.
The rendered sweeping planes are denoted by $\rgbplanes \in \mathbb{R}^{\Nchunk \times (\planescale\cdot\Ny) \times (\planescale\cdot\Nx) \times 3}$ for RGB colors and $\Tplanes \in \mathbb{R}^{\Nchunk \times (\planescale\cdot\Ny) \times (\planescale\cdot\Nx)}$ for transmittances, where $\Nchunk$ is the total number of chunks, $\Nx \times \Ny$ is the resolution of a single view of the quilt, and $\planescale$ is a resolution scaling hyperparameter. We detail how we divide the 3D volume into discrete disjoint chunks and render them for different 3D volumetric representations in Secs. \ref{sec:g2lf} and \ref{sec:v2lf}.

\subsection{Swizzle blending} \label{sec:swizzling}
We can now efficiently render the final quilt for a light field display by sub-sampling and alpha composition of these rendered sweeping planes. 
\redline{For clarity, we explicitly parameterize each quilt view by its lateral camera offset and the corresponding principal-point shift before deriving the plane-sampling coordinate.}
The quilt $\mathbf{Q} \in \mathbb{R}^{V_y \times V_x \times N_y \times N_x \times 3}$ is a 2D array of perspective views formed by laterally shifting the base camera along its horizontal ($x$) and vertical ($y$) directions at the same viewing distance, as illustrated in Fig.~\ref{fig:sweeping_planes}.
The lateral camera offsets $(\Delta x, \Delta y)$ are distributed uniformly across the viewing window at the viewing distance, and their normalized principal-point shifts \redline{$(p_x, p_y)$} are set such that the principal rays from all quilt views converge at the center of the focal plane (see Fig.~\ref{fig:sweeping_planes}).

Let each quilt view be indexed by row $i \in \{1,\dots,V_y\}$ and column $j \in \{1,\dots,V_x\}$. We define the normalized horizontal and vertical positions of each quilt view within the view grid as
\begin{equation}
s_j^x = \frac{j-1}{V_x-1} - \frac{1}{2}, \qquad
s_i^y = \frac{i-1}{V_y-1} - \frac{1}{2}.
\end{equation}
Given the display viewing angles $\phi_x$ and $\phi_y$, the corresponding camera's lateral offsets are:
\begin{equation}
\Delta x_j = 2 D_{\mathrm{focal}}\, s_j^x \tan(0.5\phi_x), \qquad
\Delta y_i = 2 D_{\mathrm{focal}}\, s_i^y \tan(0.5\phi_y),
\end{equation}
and their normalized principal-point shifts $(p_x, p_y)$ are:
\begin{equation}
p_j^x = \frac{2 s_j^x \tan(0.5\phi_x)}{\tan(0.5\theta_x)}, \qquad
p_i^y = \frac{2 s_i^y \tan(0.5\phi_y)}{\tan(0.5\theta_y)}.
\end{equation}
Given a normalized horizontal pixel coordinate $u \in [-1,1]$ in quilt view $(i,j)$, its projected horizontal coordinate $u'$ on the $k$-th sweeping plane at distance $d_k$ is
\begin{equation}
u' =
\frac{
\Delta x_j (D_{\mathrm{focal}} - d_k)/D_{\mathrm{focal}}
+
d_k \tan(0.5\theta_x)\,u
}{
(d_k - D_{\mathrm{forward}})\tan(0.5\theta_x')
}.
\label{eq:u_prime}
\end{equation}
where $\theta_x'$ denotes the modified horizontal field of view of the reference camera, as defined in Eq.~(\ref{eq:theta_prime}).
\redline{Eq.~(\ref{eq:u_prime}) maps a pixel in quilt view $(i,j)$ to the corresponding sampling location on the $k$-th sweeping plane in the reference-camera image.}
In the numerator, the first term accounts for the horizontal shift of quilt view $(i,j)$ due to the camera offset $\Delta x_j$, while the second term corresponds to the local pixel offset within that view.
We show the horizontal projection for brevity; the vertical coordinate is computed analogously using $\Delta y_i$, $\theta_y$, and $\theta_y'$.

We use Eq.~(\ref{eq:u_prime}) to project each quilt-view pixel onto the sweeping planes and sample $C$ and $T$ using either bilinear or nearest-neighbor interpolation from them. The sampled series of colors $c_k$ and transmittance values $T_m$ are blended into a final pixel color with:
\begin{equation}\label{eq:alpha_blend}
C = \sum_{k=1}^{N_{\mathrm{chunk}}} \left( \prod_{m=1}^{k-1} T_m \right) c_k .
\end{equation}
Note that the color $c_k$ here is already weighted by alpha opacity.

In the following, we introduce the representation-specific adaptations of our algorithm.

\subsection{G2LF -- 3D Gaussians to Light Field} \label{sec:g2lf}
One advantage of a primitive-based 3D volumetric representation, e.g., 3DGS which comprises of a discrete collection of 3D Gaussian primitives, is that we can have \redline{a} well-distributed rendering computational cost for each volume chunk (Fig.~\ref{fig:sweeping_planes}'s right panel) by balancing sweeping plane positions.
Specifically, we apply quantile binning so each volume chunk has similar numbers of \redline{primitives}.
For applying quantile binning with 3DGS, we \redline{select} Gaussian primitives inside the view frustum of the reference camera and compute a sorted list of the Gaussians \redline{according to} the reference camera distances.
The sorted list is partitioned into $\Nchunk$ chunks, \redline{each} with approximately the same number of elements.
The plane distance $\distplanes_k$ of the $k$-th sweeping plane is set to the median Gaussian distance of the $k$-th chunk.

\redline{A Gaussian may overlap neighboring chunks, while our implementation assigns it to a single chunk based on its center depth in the reference camera. Since sorting, chunking, and rasterization all use the same reference camera, this approximation mainly contributes to the discretization error of the sweeping-plane representation rather than introducing an additional ordering ambiguity. In practice, this error is most visible as stack-of-cards artifacts when the chunk resolution is too coarse, and can be reduced by increasing $N_{\text{chunk}}$.}

\begin{figure}
\centering
\includegraphics[width=\linewidth]{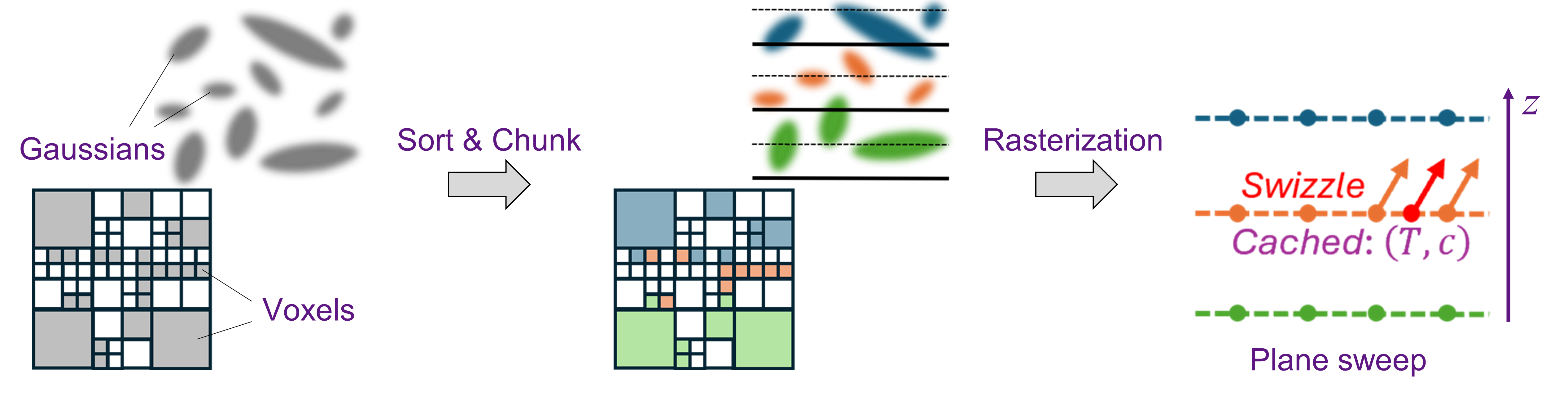}
\caption{Illustration of the G2LF and V2LF pipeline. Primitives are sorted along the $z$-axis of the reference camera and partitioned into depth chunks with similar numbers of primitives. Each chunk is rasterized onto its representative sweeping plane, producing cached color and transmittance planes, which are then composited by \textit{swizzle} blending to form the final quilt.}
\Description{A pipeline diagram of G2LF and V2LF. Scene primitives are sorted along the reference camera's depth axis and split into depth chunks with similar numbers of primitives. Each chunk is rasterized onto a representative sweeping plane to produce cached color and transmittance planes, which are then composited by swizzle blending into the final light-field quilt.}
\label{fig:g2lf}
\end{figure}

We extend the efficient CUDA rasterizer by \cite{kerbl20233d} to perform 3D tiling instead of the original 2D tiling with an additional dimension for the chunks.
A Gaussian is assigned to a tile by its patch index and chunk index.
The Gaussians in each 3D tile are sorted and rendered in parallel, which finally produces the color and transmittance forward-sweeping planes $\rgbplanes$ and $\Tplanes$.
A visualization of this rendering pipeline and its pseudocode are shown in Fig.~\ref{fig:g2lf} and Algo.~\ref{al:G2LF}, respectively.

The original 3DGS implementation uses a hard-coded antialiasing filter in pixel-size unit.
Specifically, the variance of the projected 2D Gaussian on the screen space is always dilated by $0.3$ pixels.
This causes a mismatch in the effective screen-space filter size between conventional 3DGS rendering and our sweeping-plane rendering when the plane resolution scale $P_s$ changes. (Sec.~\ref{sec:swizzling})
Intuitively, $\planescale$, alters the resolution/size of each rendered sweeping-plane, which in turn, affects the overall memory utilization of our MPI representation and its overall rendering time. 
Our solution is to align the filtering strength $f$ based on the pixel size on the focal plane.
Specifically, we use an adaptive filtering strength to align our rendering with the conventional rendering by:
\begin{equation} \label{eq:gs_prefilter_fix}
    f' = f \cdot \left(\frac{\distfocal\cdot\tan(0.5\fovx)}{(\distfocal - \distforward)\cdot\tan(0.5\fovx')} \cdot \planescale\right)^2 ~,
\end{equation}
where $f = 0.3$ is the hard-coded number in the original 3DGS implementation.
This equation yields a larger filter strength (in screen space of the reference camera) when the pixel size of the reference camera on the focal plane is smaller than the conventional one.

\subsection{V2LF -- Sparse Voxels to Light Field} \label{sec:v2lf}
The rendering of Sparse Voxels~\cite{sun2024sparse} follows the same overall principle as G2LF, since sparse voxels also form a discrete set of primitives.
We also extend the CUDA-based sparse voxel rasterizer (SVR)~\cite{sun2024sparse} in a similar way to how we extended 3DGS's rasterizer.
To handle antialiasing, SVR employs supersampling instead of pre-filtering so our filtering strength calibration Eq.~(\ref{eq:gs_prefilter_fix}) is not applicable to SVR.
However, supersampling is still too costly with a large number of chunks ($\Nchunk$).
As an alternative, we disable supersampling in SVR and apply a low-pass Gaussian filter on the sweeping planes instead, which is implemented in CUDA and performs filtering in-place without allocating extra memory.

\begin{algorithm}[t]
\caption{G2LF/V2LF \\
\( P \) - Input Gaussian or Voxel primitives \\
\( H \) - Quilt and camera parameters \\
\( Q, T \) - Output light field quilt RGB and transmittance buffer
}
\begin{algorithmic}[1]
\State \(\mathbf{d}' \leftarrow \text{CulledDepth(P)}\) \Comment{GS/Voxels z distances to camera}
\State \(\distplanes \leftarrow \text{FindQuantile}(\mathbf{d}')\) \Comment{\(\Nchunk\) plane distances}
\State \( \{P'_k\} \leftarrow \text{Sort\_and\_Chunk}(P, \distplanes) \) \Comment{Sort \& Chunk GS/Voxels}
\State \textbf{Parallel for all} \( k \) \textbf{do} \Comment{Rasterize chunks to midplane}
    \State \hspace{1.5em} \( \Tplanes_k, \, \rgbplanes_k \leftarrow \text{Rasterize}(H, P'_k) \) \Comment{Cache grid values}
\State \( Q, T \leftarrow \text{Initialize\_buffers()} \) 
\For{\( k \) in \( 1 ... \Nchunk \)} \Comment{Swizzle blending}
    \State \(U \leftarrow \text{Quilt2PlaneCoordinate}(H, \distplanes_k)\) \Comment{Eq.~(\ref{eq:u_prime})}
    \State \(c \leftarrow \text{interpolate}(U, \rgbplanes_k)\)
    \State \(t \leftarrow \text{interpolate}(U, \Tplanes_k)\)
    \State \(Q, T \leftarrow (Q + T \odot c), (T \odot t)\) \Comment{Eq.~(\ref{eq:alpha_blend}), Volume rendering}
\EndFor
\end{algorithmic}
\label{al:G2LF}
\end{algorithm}

\begin{table}[h]
\centering
\caption{Light field quilt rendering results comparison on the Mip-NeRF360 dataset using various algorithms and their ablations. The results are averaged on 9 scenes where we sample 4 viewpoints to render light field quilts for each scene (quilt resolution set to \(\vx{=}45, \vy{=}1, \Nx{=}\Ny{=}512\)). 3DGS~\cite{kerbl20233d} and SVR~\cite{sun2024sparse} are used for the baselines whose rendered quilts are served as the ground truth for G2LF and V2LF, respectively. The FPS indicate the rendering speed of the entire quilts measured on a NVIDIA RTX 5090\redline{, and memory usage (MEM) is reported in GB.}
}
{{
\begin{tabular}{@{}lcc ccc@{}}
\toprule
\textbf{Method} & \textbf{FPS↑} & \textbf{{MEM↓}} &
\textbf{LPIPS↓} & \textbf{PSNR↑} & \textbf{SSIM↑} \\
\midrule
\textbf{3DGS}             &  5.9  & {5.8} &  \multicolumn{3}{c}{serve as ground truth} \\
\textbf{G2LF ($\Nchunk{=}64$})  &  127 & {6.3} & 0.311 & 24.89 & 0.749 \\
\textbf{G2LF ($\Nchunk{=}64, \planescale{=}2$})  & 109 & {7.5} & 0.129 & 26.96 & 0.820 \\
\textbf{G2LF ($\Nchunk{=}512$}) &   50 & {8.5} & 0.277 & 26.89 & 0.831 \\
\textbf{G2LF ($\Nchunk{=}512$, Bi.)} &  24 & {8.5} & 0.156 & 28.78 & 0.875  \\
\textbf{G2LF ($\Nchunk{=}512, \planescale{=}2$}) &  34 & {16.2} & 0.084 & 30.71 & 0.924  \\
\textbf{G2LF ($\Nchunk{=}512, \planescale{=}2$, Bi.}) & 19 & {16.2} & 0.065 & 33.31 & 0.950   \\
\textbf{G2LF ($\Nchunk{=}512, \planescale{=}3$, Bi.}) & 15 & {29.0} & 0.053 & 34.77 & 0.958   \\
\cmidrule{2-6}
\textbf{SVR}              &  8.2  & {4.0} &  \multicolumn{3}{c}{serve as ground truth} \\
\textbf{V2LF ($\Nchunk{=}64$})  &  133 & {4.4} & 0.344 & 23.63 & 0.673 \\
\textbf{V2LF ($\Nchunk{=}64, \planescale{=}2$})  & 104 & {4.9} & 0.153 & 26.97 & 0.793 \\
\textbf{V2LF ($\Nchunk{=}512$}) &   48 & {5.7} & 0.302 & 25.57 & 0.767 \\
\textbf{V2LF ($\Nchunk{=}512$, Bi.)} &  26 & {5.7} & 0.182 & 28.51 & 0.845  \\
\textbf{V2LF ($\Nchunk{=}512, \planescale{=}2$}) &  26 & {10.2} & 0.099 & 31.06 & 0.908  \\
\textbf{V2LF ($\Nchunk{=}512, \planescale{=}2$, Bi.}) & 17 & {10.2} & 0.072 & 33.87 & 0.938   \\
\textbf{V2LF ($\Nchunk{=}512, \planescale{=}3$, Bi.}) & 11 & {17.8} & 0.055 & 35.54 & 0.950   \\
\bottomrule
\end{tabular}
}}
\label{tab:quilts_views}
\end{table}
\section{Results}

\subsection{Evaluation}
\paragraph{Dataset.} We use the four indoor and five outdoor unbounded real-world scenes from the Mip-NeRF360 dataset for evaluation. For training we use all the training set cameras and then sample four viewpoints from the test set for evaluating the rendering quality.

\paragraph{Baseline models reproduction.}
We used 3DGS~\cite{kerbl20233d} and SVR~\cite{sun2024sparse} as the baseline models. We directly train with their default hyperparameters on the Mip-NeRF360 dataset. The standard perspective novel-view results are (LPIPS 0.215, PSNR 27.53, SSIM 0.816) for 3DGS and (LPIPS 0.186, PSNR 27.36, SSIM 0.822) for SVR averaged across all scenes. The proposed G2LF and V2LF algorithms are evaluated against these trained baseline models.

\paragraph{Results on light field quilt.}
In Fig. \ref{fig:teaser} (left), the render FPS obtained for varying number of views is presented using different algorithms. Our G2LF and V2LF algorithms (solid lines) achieve real-time performance (>60 FPS) even for over 90 views, enabling interactive 3D visualizations of scenes on light field displays.

Table \ref{tab:quilts_views} presents a quantitative comparison of light field quilts (\(V_x{=}45, V_y{=}1, N_x{=}N_y{=}512\)) rendered using various algorithms and their ablative variants. As we do not have the ground truth images for the quilts, we use the rendering results for the quilts from the conventional method which individually renders multiple views to serve as the ground truth for G2LF and V2LF. 
These results demonstrate that G2LF and V2LF can faithfully render light field quilts while substantially accelerating rendering compared with their conventional counterparts, 3DGS and SVR. The number of chunks $N_{\text{chunk}}$, the sweeping-plane resolution scale $P_s$, and the interpolation method (nearest-neighbor by default; `Bi.' for bilinear) are the main parameters controlling the speed--quality trade-off. For example, with $N_{\text{chunk}} = 64$, G2LF and V2LF achieve 127 FPS and 133 FPS, respectively, compared with 5.9 FPS and 8.2 FPS for conventional 3DGS and SVR quilt rendering, while maintaining reasonable image fidelity.

\begin{figure}[t]
\centering
\includegraphics[width=\linewidth]{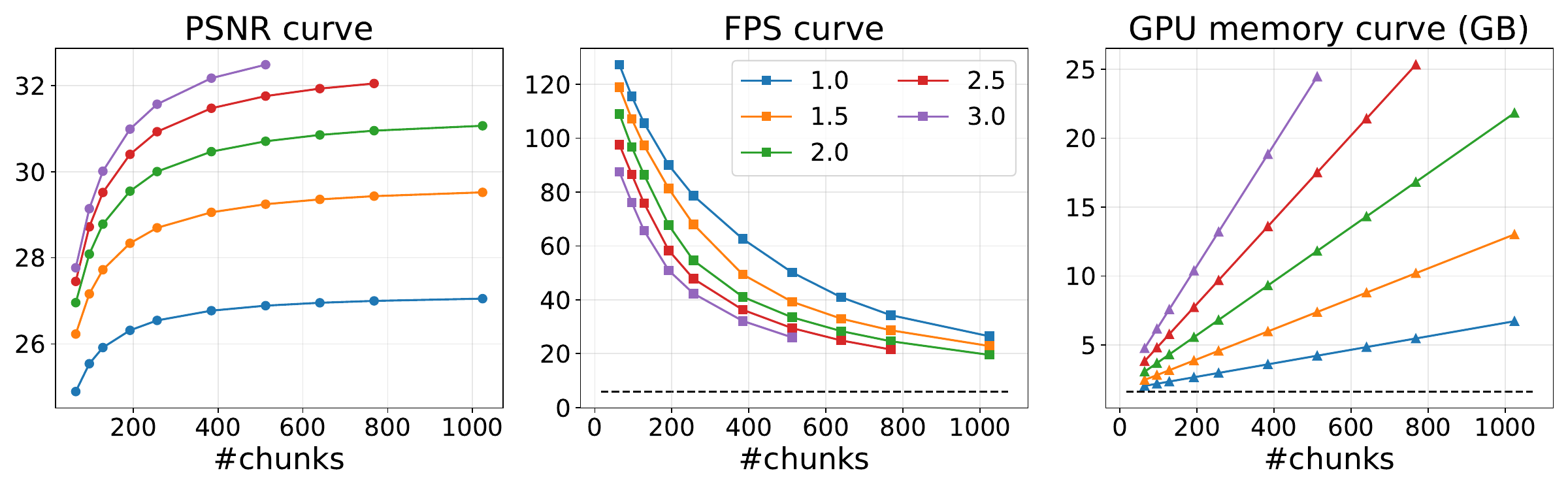}
\caption{
Performance curves of G2LF under various $\Nchunk$ (x-axis) and $\planescale$ (colors) values.
Some curves have fewer data points due to VRAM limits at those configurations.
In comparison, the black dash lines are the FPS and required memory for conventional 3DGS rendering.
More planes and higher plane resolution both consistently improve quality at the cost of \redline{increasing} compute and memory usage.
The quality (PSNR) improvement is about saturated at around $\Nchunk{=}400$.
Higher $\planescale$ values keep on improving the quality, but at the cost of greater memory usage.
}
\Description{Performance curves for G2LF plotting quality, frame rate, and memory against the number of depth chunks on the x-axis, with colored curves for different sweeping-plane resolution scales; black dashed lines mark conventional 3DGS. Quality improves with more chunks and higher plane resolution but saturates near 400 chunks, while memory and compute cost rise.}
\label{fig:perf_g2lf}
\end{figure}

Increasing the depth resolution ($N_{\text{chunk}}$) and plane resolution scale ($P_s$) brings the rendered quilts closer to the baseline quality, at the cost of higher computation and memory usage. As shown in Fig.~~\ref{fig:perf_g2lf}, both parameters consistently improve reconstruction quality, while Fig.~\ref{fig:render} confirms that the resulting perspective views remain visually accurate in real-time. In particular, increasing $P_s$ reduces interpolation errors during the swizzle operation.
Performance curves in Fig.~\ref{fig:perf_g2lf} show many ablative variations of G2LF with different number of chunks and plane resolution scales, both of which consistently mitigate image-quality differences from the slower conventional rendering approach. We also find that using 8-bit unsigned integer values to store $C$ and $T$ leads to similar blending quality compared to using 32-bit floating-point values (see Fig.~\ref{fig:perf_uint8}). We therefore use 8 bits by default, which substantially reduces memory usage and improves rendering speed.

\begin{figure}[t]
\centering
\includegraphics[width=\linewidth]{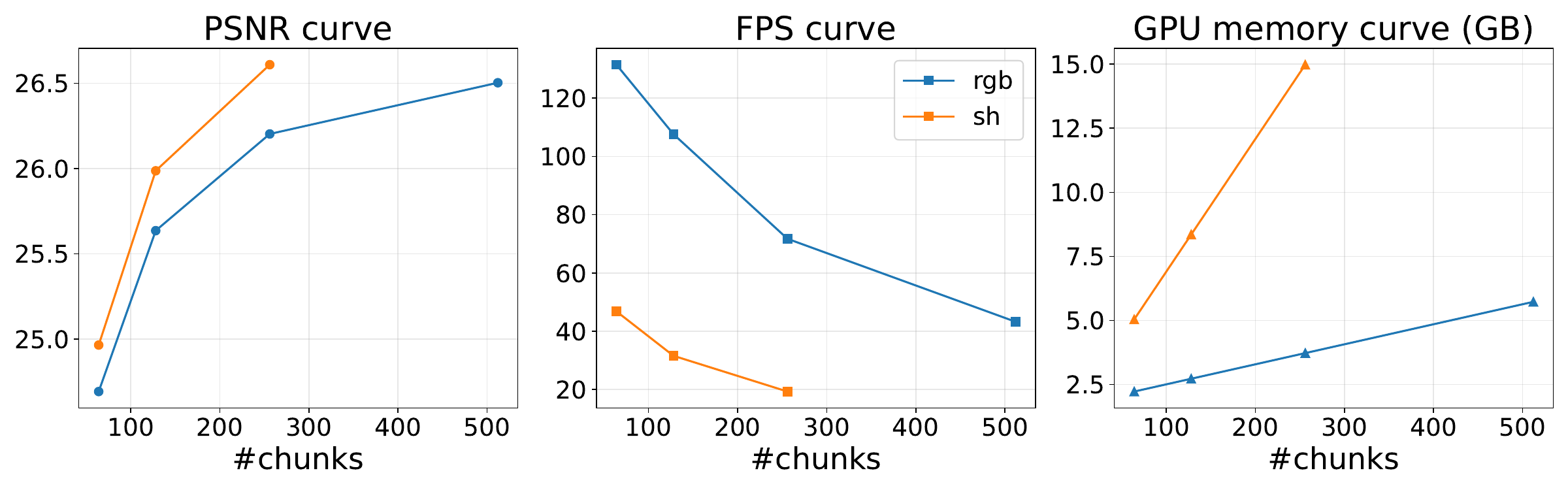}
\caption{Comparison of \textit{swizzle} blending with RGB color or spherical harmonic (SH) for more accurate view-dependent colors. The quality improvement (PSNR) is relatively minor compared to using more chunks or higher plane resolution (Fig.~\ref{fig:perf_g2lf}). However, with spherical harmonics the drop in FPS and the increase in memory usage is significant. \redline{We therefore use RGB-alpha swizzle blending by default, accepting reduced view-dependent accuracy for off-center views.}}
\Description{Plots comparing swizzle blending with plain RGB color versus spherical-harmonics color. Spherical harmonics gives only a small PSNR gain while substantially lowering frame rate and raising memory, so RGB-alpha blending is used by default.}
\label{fig:perf_shs}
\end{figure}

\begin{figure}[t]
\centering
\includegraphics[width=\linewidth]{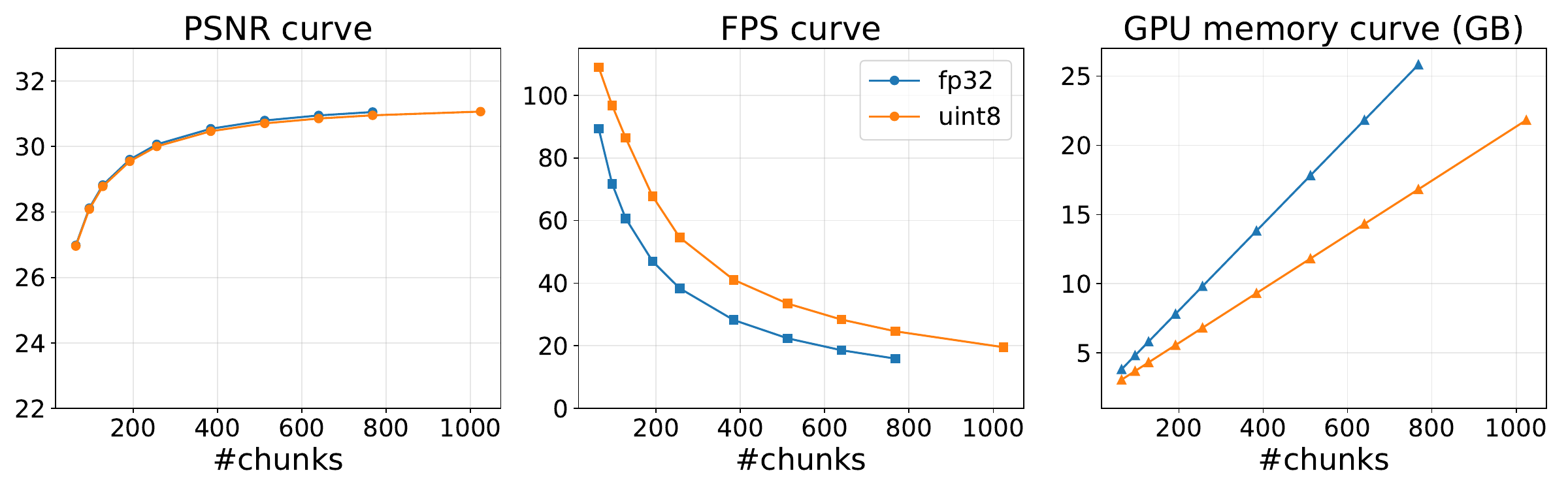}
\caption{Comparison of using uint8 and fp32 to store the sweeping planes of G2LF ($\planescale{=}2$). The PSNR is almost identical while the run time and memory usage is largely reduced with uint8. Switching to uint8 also results in being able to use \redline{a higher number of} chunks without running out of memory, versus fp32.}
\Description{Plots comparing 8-bit integer versus 32-bit floating-point storage of the sweeping planes. PSNR is nearly identical for both, while the 8-bit version markedly reduces runtime and memory and allows more chunks before exhausting memory.}
\label{fig:perf_uint8}
\end{figure}

\begin{figure}[t]
\centering
\includegraphics[width=\linewidth]{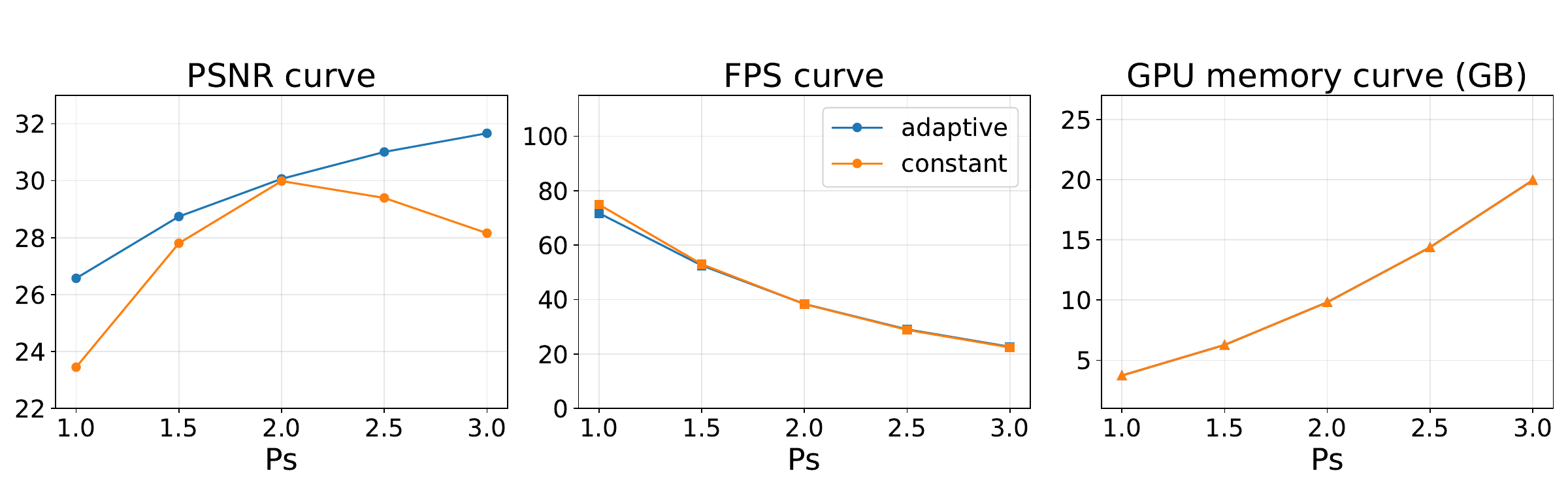}
\caption{Ablation of adaptive versus constant anti-aliasing filter strengths in G2LF ($N_{\text{chunk}}=256$). 
Using a constant filter strength leads to a screen-space filter mismatch between conventional 3DGS rendering and our sweeping-plane rendering when the plane resolution scale $\planescale$ changes. 
In contrast, the adaptive filter maintains consistent quality improvements with increasing $\planescale$, while having nearly identical FPS and memory usage.}
\Description{Plots comparing adaptive versus constant anti-aliasing filter strength as the sweeping-plane resolution scale changes. The constant filter suffers a screen-space mismatch that degrades quality, whereas the adaptive filter keeps consistent quality gains at similar frame rate and memory.}
\label{fig:perf_adapt}
\end{figure}

\begin{figure}[t]
\centering
\includegraphics[width=\linewidth]{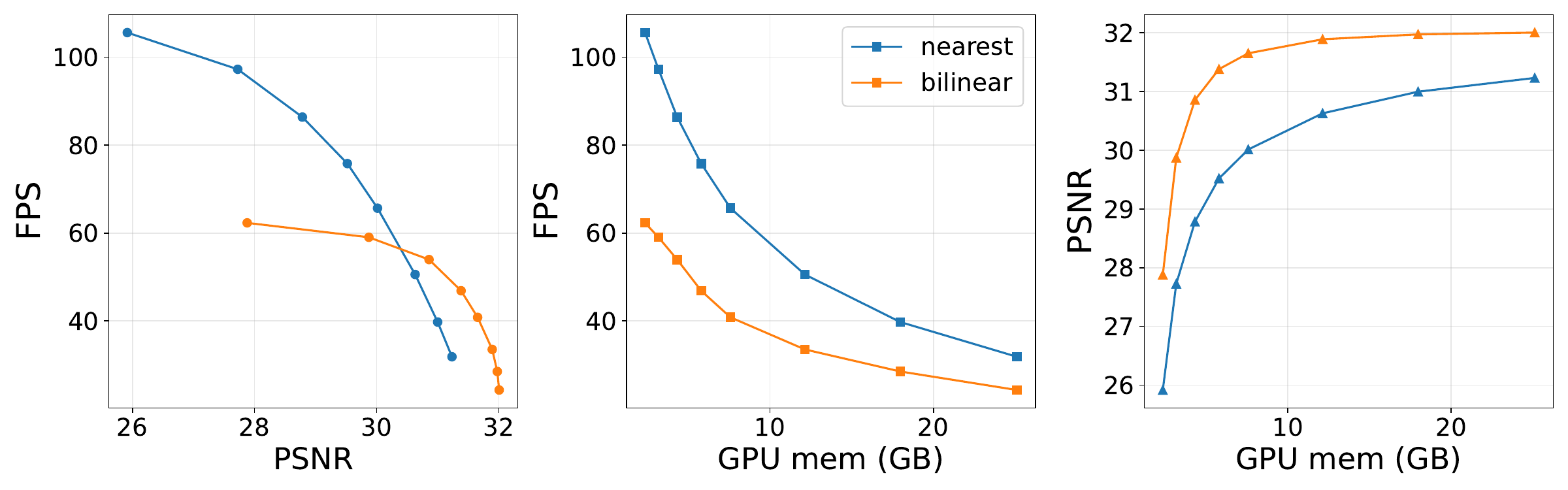}
\caption{Comparing bilinear and nearest-neighbor interpolation with \textit{swizzle} blending with $\Nchunk{=}64$ and $\planescale \in [1, 6]$. As expected, bilinear interpolation achieves better quality but with slower FPS under the same amount of GPU memory usage comparing to nearest-neighbor interpolation. There is a crossover point in left figure at around 30.5 PSNR. To render in lower quality, nearest-neighbor is faster while bilinear becomes faster for achieving higher PSNR. This is because nearest-neighbor interpolation needs higher $\planescale$ to achieve the same quality as bilinear, which becomes the dominant factor for computation cost with high $\planescale$. Theoretically, both versions should achieve the same PSNR with infinite $\planescale$.}
\Description{Plots comparing bilinear versus nearest-neighbor interpolation in swizzle blending across plane-resolution scales. Bilinear gives higher quality at lower frame rate; nearest-neighbor is faster at lower quality, with the curves crossing near 30.5 dB PSNR at equal GPU memory.}
\label{fig:perf_bilinear}
\end{figure}

\begin{figure}[t]
\centering
\includegraphics[width=0.9\linewidth]{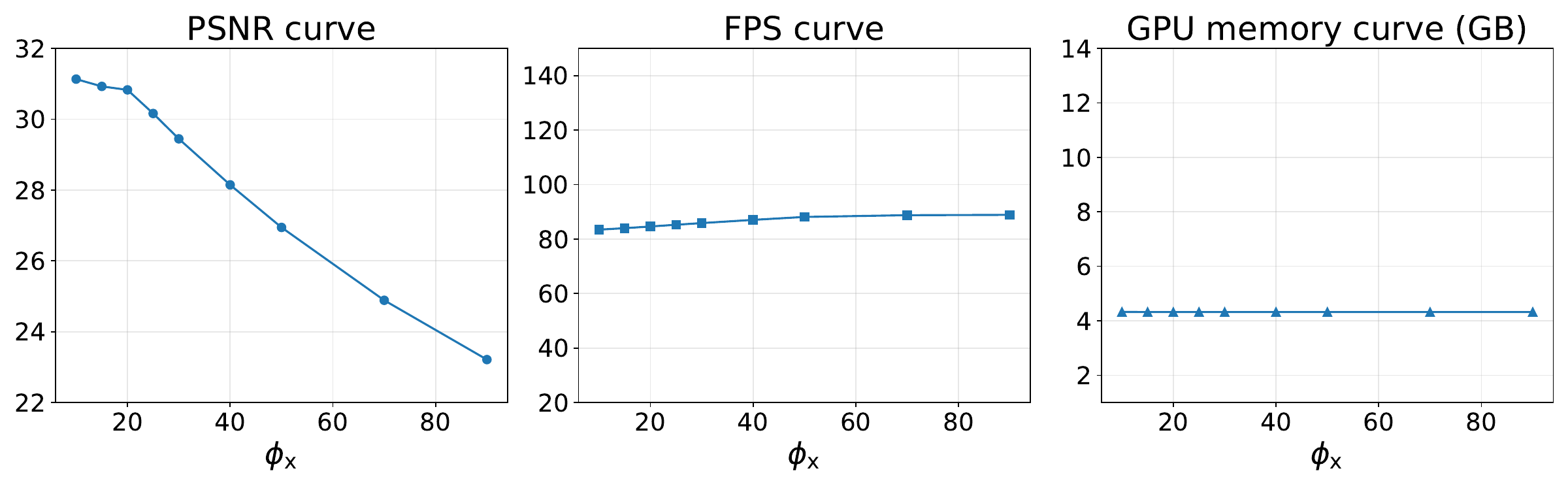}
\hfill
\includegraphics[width=0.9\linewidth]{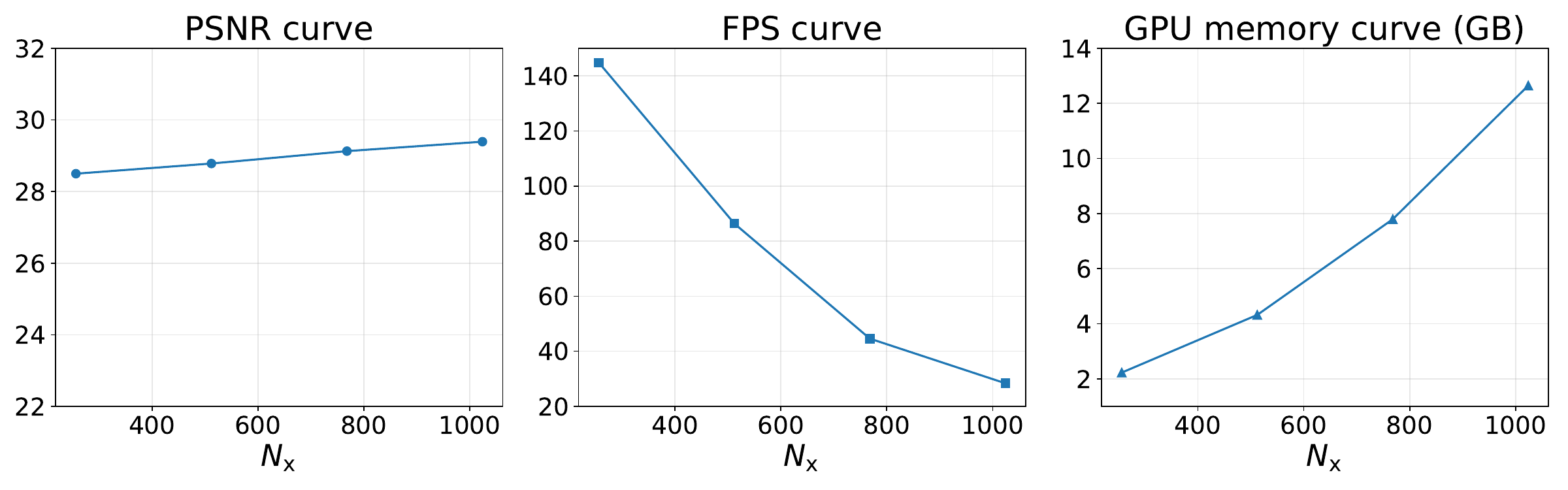}
\hfill
\includegraphics[width=0.9\linewidth]{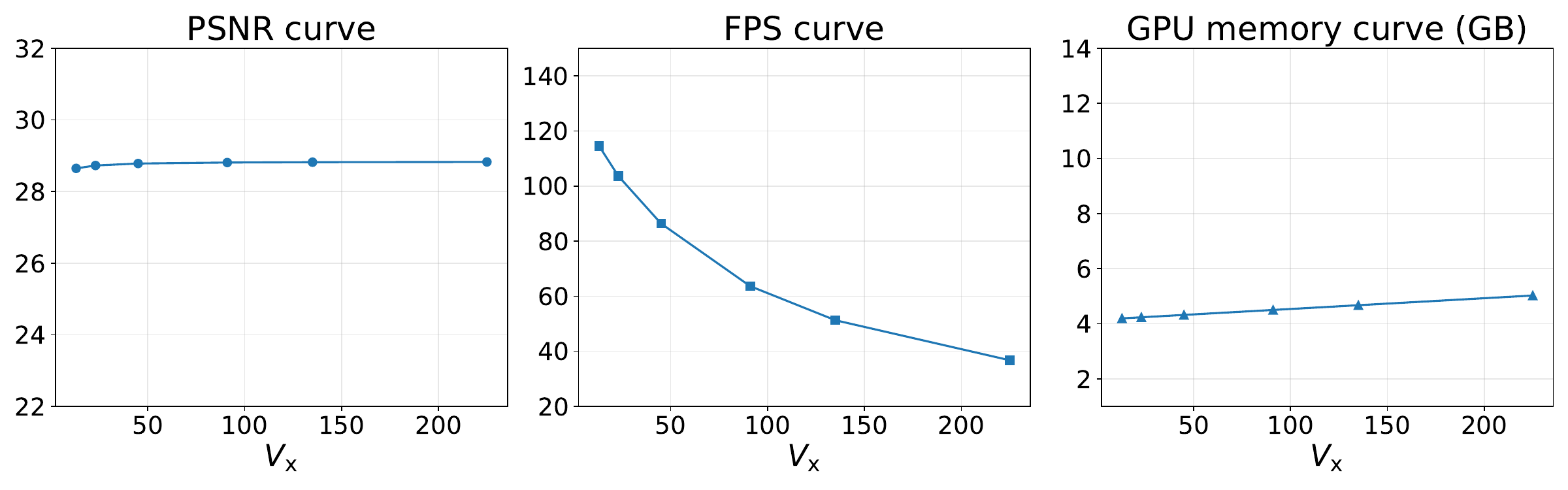}
\caption{Results of G2LF under different quilt settings. The default setting is $\planescale{=}2, \Nchunk{=}128, \viewfovx{=}35, \viewfovy{=}0, \Nx{=}\Ny{=}512, \vx{=}45, \vy{=}1$, with nearest-neighbor interpolation and uint8 sweeping planes. 
Top: increasing the viewing angle slightly reduces quality, while FPS and memory usage remain nearly unchanged. 
Middle: increasing the per-view quilt resolution increases FPS cost and memory usage approximately proportionally, while the quality improvement remains modest. 
Bottom: increasing the number of quilt views raises runtime and memory usage, while PSNR remains largely similar within the same viewing range.}
\Description{Three stacked plots of how G2LF responds to quilt settings. Top: a larger viewing angle slightly lowers quality with little change in speed or memory. Middle: higher per-view resolution raises frame-time and memory roughly proportionally with modest quality gains. Bottom: more quilt views raise runtime and memory while PSNR stays largely constant.}
\label{fig:perf_quilt}
\end{figure}

We show comprehensive ablation studies of \redline{the various} individual components of G2LF and discuss their results in the captions---ablating spherical harmonic in Fig.~\ref{fig:perf_shs}, comparing uint8 and fp32 in Fig.~\ref{fig:perf_uint8}, showing the effect of the adaptive anti-aliasing filter in Fig.~\ref{fig:perf_adapt}, comparing different interpolation algorithms in Fig.~\ref{fig:perf_bilinear}, and different quilts setup in Fig.~\ref{fig:perf_quilt}.

\subsection{Captured Results}
We captured the displayed light field results for several algorithm variants using a Looking Glass Go display \cite{lookingglass} and \redline{through} a Google Pixel 9 Pro (f/2.8, 1/50s, ISO 100) \cite{google}, as shown in Fig.~\ref{fig:capture}.
The display was mounted on a motorized rotational stage, and viewpoints were uniformly sampled across the horizontal range ($-18.5^\circ$ to $+17.5^\circ$) at 1.5$^\circ$ intervals. The brightness of the captured images was adjusted linearly for visualization. The video results confirm that our method enables high-quality 3D reconstruction with accurate angular consistency across views (see supplementary video). In practice, artifacts visible in individual rendered views are less noticeable during actual use, as multiple views blend perceptually within the viewer’s pupil. See our supplementary materials for more captured results.

\subsection{Real-Time Interactive 3D Visualization Demo}\label{sec:interactive}

\begin{figure}[t!]
\centering
\includegraphics[width=0.7\linewidth]{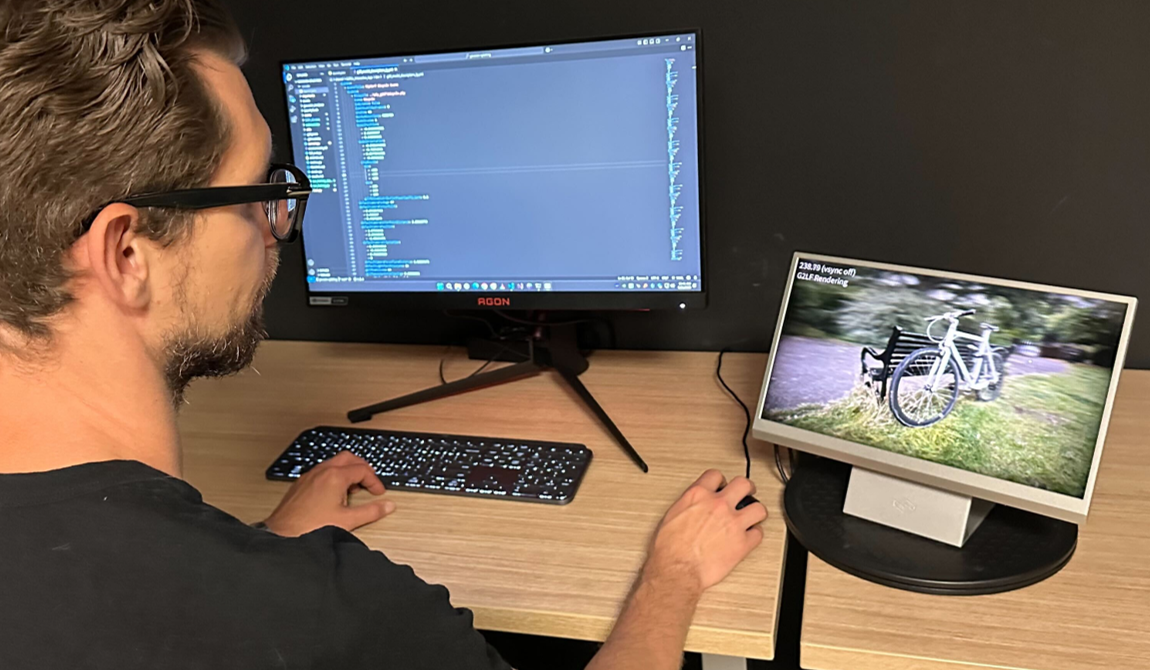}
\caption{Real-time interactive 3D radiance field demonstration on a Looking Glass light field display, enabling users to dynamically change viewpoints and visualize radiance fields in a 3D space \cite{kim2025play4d}.}
\Description{A photograph of the real-time interactive demo: a Looking Glass light field display driven by a desktop with a single NVIDIA RTX 5090 shows a 3D Gaussian scene that viewers can navigate interactively, perceiving depth through binocular parallax.}
\label{fig:demo}
\end{figure}

\begin{figure*}[h]
\centering
\includegraphics[width=0.9\textwidth]{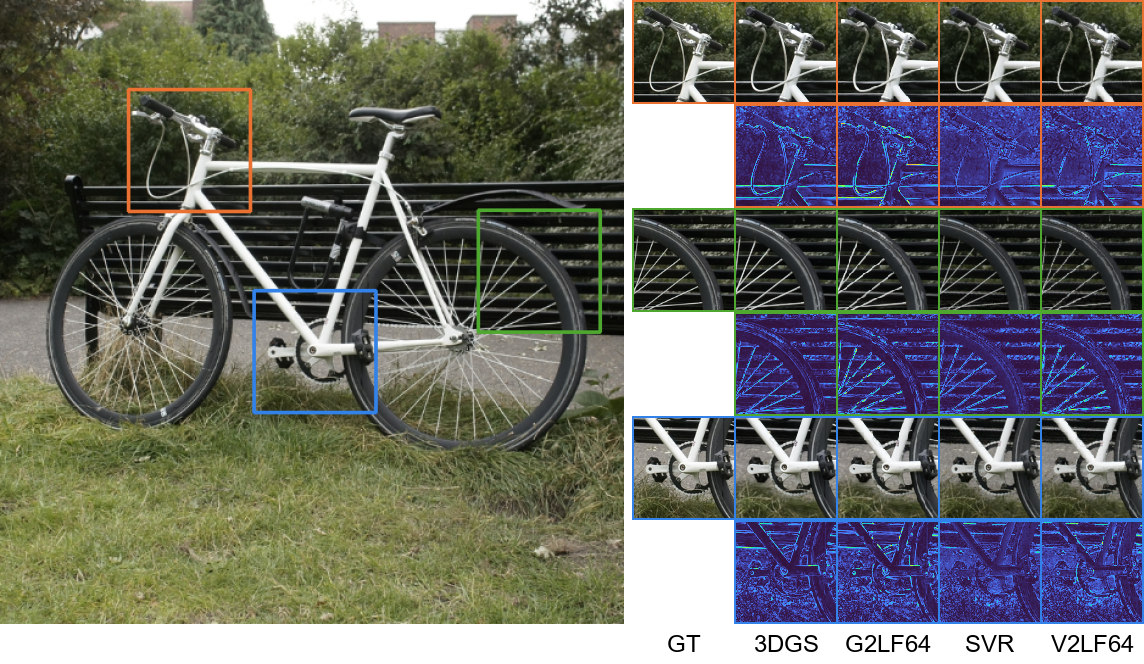}
\includegraphics[width=\textwidth]{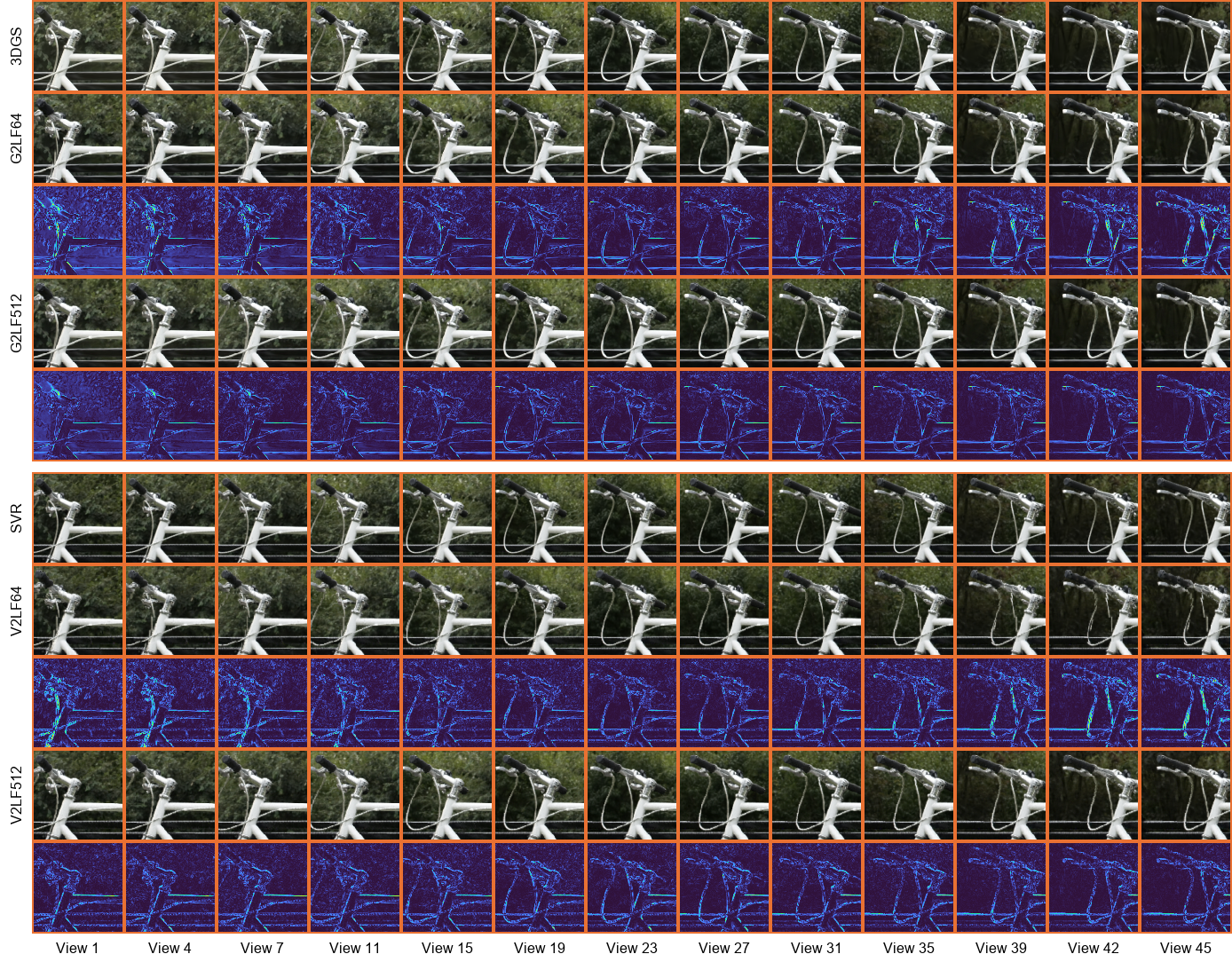}
\caption{Evaluation on the rendered \textsc{Bicycle} scene from the MipNeRF-360 dataset \cite{barron2022mip}. Top: visual comparisons between the ground truth, the baseline models, and our proposed methods, \redline{with error maps highlighting per-pixel differences from the reference images}. Bottom: rendered light field quilts across viewing angles, \redline{where the accompanying error maps show view-dependent rendering differences}. The numbers appended to G2LF and V2LF denote the $N_{\text{chunk}}$.}
\Description{Evaluation on the Bicycle scene from MipNeRF-360. The top row compares the ground truth, baseline models, and our methods, with per-pixel error maps highlighting differences from the reference. The bottom row shows the rendered light-field quilts across viewing angles with error maps of view-dependent differences; numbers after G2LF and V2LF indicate the chunk count.}
\label{fig:render}
\end{figure*}

We demonstrate the application of our method for dynamic, interactive 3D visualization on a light field display \cite{kim2025play4d}. Fig.~\ref{fig:demo} showcases the demo running on a desktop equipped with a single NVIDIA RTX 5090 and a Looking Glass 16" Light Field Display. In comparison to a traditional 2D display, our real-time G2LF implementation allows users to perceive depth directly through binocular cues, providing a more intuitive and immersive interaction with 3DGS content.
Our interactive application is implemented in OpenGL/C++ to make use of texture filtering and blending in hardware during the \textit{swizzle} operation.
We use 8-bit RGBA framebuffers for both sweeping plane buffers and the 45-views 4K x 4K quilt. 
Our OpenGL implementation mostly follows our description in Alg. \ref{al:G2LF}. One difference is that we switch the order of quantile computation and sorting to sort all primitives in one pass upfront.
From the rendered quilt we apply interlacing provided by the LookingGlass Bridge SDK before sending the frame to the swapchain. 

\begin{figure*}[t!]
\centering
\includegraphics[width=0.95\linewidth]{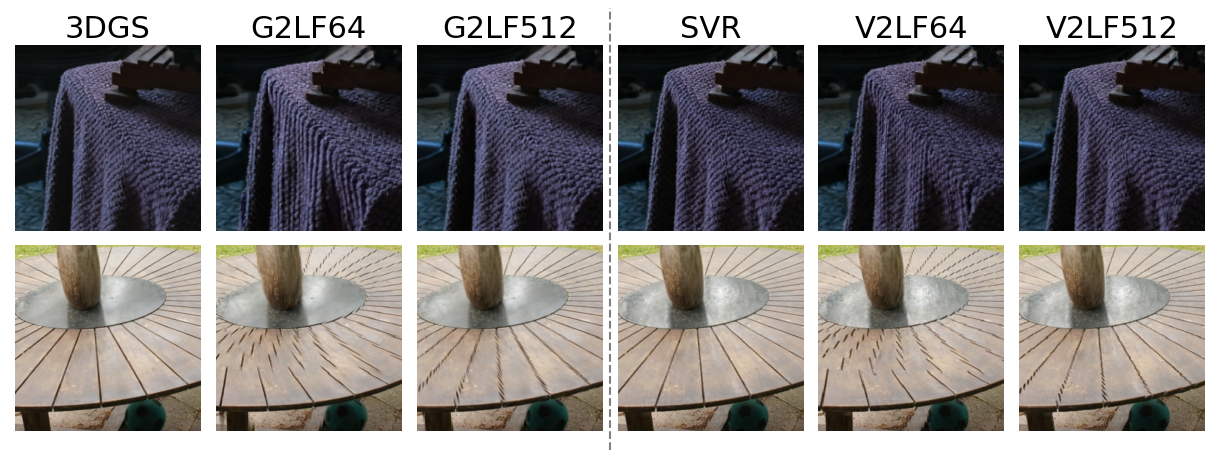}
\caption{Close-up comparison of the \textsc{bonsai} and \textsc{garden} scene from the MipNeRF-360 dataset \cite{barron2022mip} at the (top) left-most (View 45) and (bottom) right-most (View 1) viewing angle. Increasing the number of chunks reduces depth discontinuity errors.}
\Description{Close-up comparisons on the Bonsai and Garden scenes at the left-most (view 45, top) and right-most (view 1, bottom) viewing angles. Increasing the number of depth chunks progressively reduces depth-discontinuity artifacts (the stack-of-cards effect).}
\label{fig:ablation} 
\end{figure*}

\begin{figure*}[h]
\centering
\includegraphics[width=0.98\textwidth]{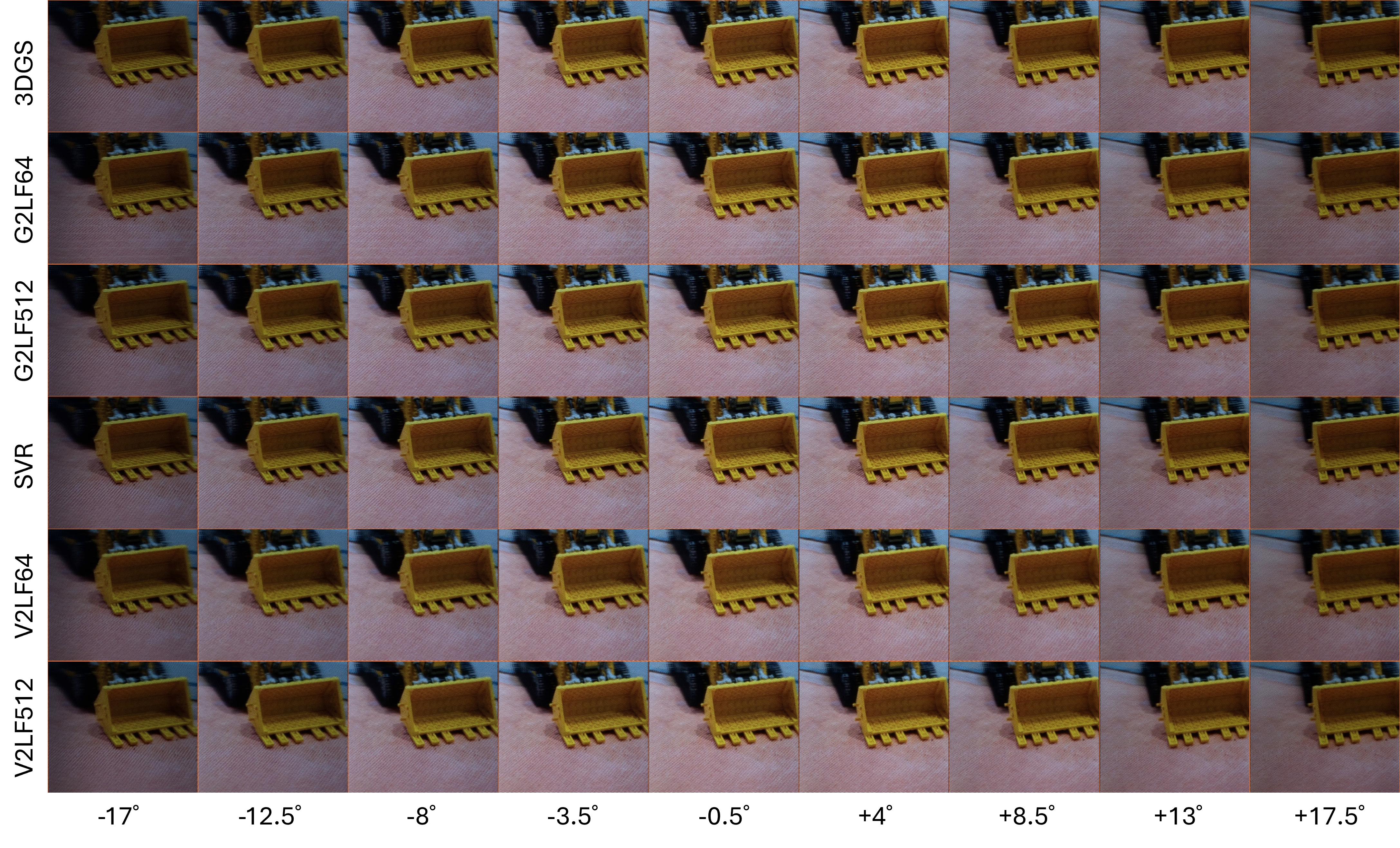}
\caption{Captured results of the \textsc{Kitchen} scene from the MipNeRF-360 dataset \cite{barron2022mip} for multiple algorithm variants on a Looking Glass Go display. All variants produce geometrically consistent light-field rendering and faithful 3D reconstruction under real display conditions. In practice, artifacts visible in individual rendered views are less noticeable on the display, as multiple views perceptually blend within the viewer's pupil. More captured results are provided in the supplementary materials. }
\Description{Photographs of the Kitchen scene from MipNeRF-360 captured from a Looking Glass Go display for several algorithm variants, showing geometrically consistent light-field rendering and faithful 3D reconstruction under real display conditions.}
\label{fig:capture}
\end{figure*}

Our optimized OpenGL implementation achieves comparable image quality to the Python/CUDA implementation presented in Tab.~\ref{tab:quilts_views}, but achieves better computational performance due to higher GPU utilization. Compared to the Python/CUDA baseline, the speedup is approximately $3.6\times$ on average across tested scenes ($\Nchunk{=}64, \planescale{=}1$, measured on NVIDIA RTX 5090).
Our GL implementation scales well across varying number of chunks. In our tests $\Nchunk{=}64$ to $\Nchunk{=}128$ provide a sweet spot in terms of performance. Although higher numbers of chunks result in slightly higher image quality metrics the quality increase diminishes when perceived on the light field display (see supplementary video). Lower chunk numbers ($\Nchunk{<}32$) can result in noticeable discrete depth layers for surfaces close to the camera \redline{("stack-of-cards effect" as shown in Fig.~\ref{fig:ablation})} and should be avoided.

\section{Discussion}
\paragraph{Framework Versatility}
Our unified framework supports a wide range of radiance field representations, including all explicit formats like 3D Gaussians and Sparse Voxels, without the need for retraining. \redline{Besides static 3D scenes this allows our method to directly support 4D immersive content as well.} Further, one can apply our rendering algorithm to implicit representations like NeRF variants. To test this, we used the coarse network in NeRF to render a roughly estimated depth map from the base camera view first and quantile the depth points to determine the chunk positions. Despite the effort, we still observe apparent degradation in quality of N2LF comparing to using the conventional quilt rendering with NeRF (LPIPS 0.379, PSNR 23.64, SSIM 0.822 with Nerfacto~\cite{tancik2023nerfstudio} baseline), mostly because of  poor importance sampling. Designing a specific sampling method to address this is out of our current scope and we leave it for future work.

Our algorithm is compatible with all types of light field displays, from 3D displays with head-tracked rendering to integral imaging systems offering vertical parallax. Unlike methods limited to a fixed set of view directions, our slice-based approach enables fast reconstruction from arbitrary viewpoints within the supported viewing angle.

\paragraph{Limitations and Future Work.}
One of the main limitations of our approach is the increased memory usage due to the intermediate slice-based representation, which is more demanding than direct rendering methods. Future work may explore reducing this memory footprint, for example by employing hardware-accelerated compression or more compact data layouts.

Additionally, blending performance is inherently tied to the number and resolution of slice planes. While more slices enable finer volumetric detail, they also incur higher computational cost. \redline{If the number of chunks is too small, the discretized slice representation can introduce stack-of-cards artifacts, particularly in scenes with large depth variation or geometry close to the viewer. This scene-dependent trade-off can be mitigated by increasing the number of chunks at the cost of additional memory and computation.}

\redline{Our method also trades off view-dependent appearance for rendering efficiency. Because scene appearance is baked into shared sweeping planes, strongly view-dependent effects such as specular highlights and reflections may be less accurate for off-center views than in per-view rendering. However, our spherical-harmonic ablation shows that preserving view-dependent color provides only minor quality improvement while substantially increasing memory usage and reducing rendering speed (Fig.~6). We therefore use RGB sweeping planes by default, which offers a better practical trade-off for real-time light field display rendering.}

Since the ultimate output is constrained by the characteristics of physical light field displays (e.g., limited angular or depth resolution), future optimization could benefit from adapting the slice structure to better match the display-specific capabilities and perceptual requirements.

\bibliographystyle{ACM-Reference-Format}
\bibliography{sample-base}

\end{document}